\documentstyle[]{article}
\begin{document}
 
\title{
Some remarks on the angular momenta of galaxies,  their clusters
and superclusters}

\author{
W{\l}odzimierz God{\l}owski${^1}$
Marek Szyd{\l}owski ${^1}$
Piotr Flin$^{2}$
}
 
\maketitle
 
1. Astronomical Observatory of the Jagiellonian University, 30-244
Krakow, ul. Orla 171, Poland  e-mail: godlows@oa.uj.edu.pl
2.Pedagogocal University, Institute of Physics, 25-406 Kielce ul Swietokrzyska 15 
Poland

\section*{Abstract}

We discuss the relation between angular momenta and masses of galaxy
structures base on the  Li model of the universe with global rotation.
In our previous paper (God{\l}owski et al 2002) it was shown that  
the model predicts the presence of a minimum in this relation. In the present
paper we discuss observational evidence allowing us to verify this relation.
We check  these theoretical
predictions  analysing Tully's galaxy grups.
We find null angular momentum $J=0$ for the masses corresponding to mass
of galaxy grups  and non-vanishing
angular momenta for other galaxies structures. 
The  comparing of alignment in different galactic structures are
consistent with obtained theoretical relation $J(M)$ if we interpret the
groving alignment as the  increasing angular momenta  of galaxies in the large
scale.

\section{Introduction}

The  empirical relation between the angular momentum and mass
of celestial bodies was investigated for a long time.
\cite{Wesson79, Wesson83, Carrasco82, Brosche86}.
Usually this relation is presented as $J\sim M^{5/3}$.
There are several methods of explanation single relation $J \propto M^{5/3}$.
Muradyan  explained this relation in terms of the Ambarzumian's
superdense cosmogony  \cite{Muradyan75}. Mackrossan  involved
termodynamical consideration  for its explanation \cite{Mackrossan87}.
Wesson (1983) argued that this is a consequence of self similarity
of Newtonian problem applied to rotating gravitionally bound systems 
\cite{Wesson83}. The considerations connected with this
relation were involved in showing its possible important role
in the unification of gravitation and particle physics
\cite{Wesson81}, as well as in constructing the new universal constant
\cite{DeSabbata84}.
Later  on Catalan and Theuns explained the relation $J\sim M^{5/3}$ in
the tidial torque model \cite{Catelan96}. Sistero incorporated the 
rotational velocity of the Universe \cite{Sistero83}.
A similar approach was presented by Carrasco, Roth {\&} Serrano, who
explained this relation as a consequence of mechanical equilibrium
between the gravitational and rotational energy\cite{Carrasco82}.
 
Already it  was proposed by Li more general relation $J(M)$ \cite{Li98}. 
It predict different relation for low magnitude of masses than simple 
$J \sim M^{5/3}$. Only for high
masses simply relation between the angular momentum and mass of celestial
bodies $J\sim M^{5/3}$ is satisfies. The crucial point of his  paper \cite{Li98} 
is the explanation  this relation for galaxies as a result of the influence
of the global rotation of the Universe on their formation.

The problem of the rotation of the whole Universe  attracted the attention
of several scientists since \cite{Birch82}.
It was shown that the reported value of rotation is too big when compared with
the CMB anisotropy. Silk  \cite{Silk70} pointed out that the dynamical effects
of a general rotation of the Universe are  presently unimportant, contrary
to the early Universe for which angular velocity $\Omega \ge 10^{-13}
rad/yr$. He stressed that now the period of rotation must be greater than
the Hubble time, which is a simple consequence of the CMB isotropy.
Barrow, Juszkiewicz and Sonoda also addressed this question \cite{Barrow85}. 
They showed that the cosmic vorticity depends strongly on
 the cosmological models and assumptions connected with linearisation
of homogeneous, anistropic cosmological models over the isotropic
Friedmann Universe. For a flat universe the obtained the value
$\omega/H_0= 2 \cdot 10^{-5}$.

It is interesting to confront the generalized relation obtained from the Li 
model with the observations. In the previous paper \cite{Godlowski03}
it was shown that the Li relation between the angular momentum and the mass 
of the celestial bodies posses a minimum. In the present paper we discuss 
observational evidence allowing us to verify this relation.
 The angular momentum of a structure is the 
sum of spins of their components and its own angular momentum.
There is no clear empirical evidence that galaxy groups and cluster rotate.
Therefore,  the study of galaxy alignment (i.e. ordering of galaxy
axes or its normal vectors to galaxy planes) in a large structure can be
useful for testing, if the primordial  correlation of galaxy spins can be
still observed in present-day structures.  in order to check Li relation we 
need to  calculate alignment for different size structures. 

The observations indicate that different galactic structures have 
different total angular momentum. We consider pair of galaxies, compact 
galaxy groups, groups of galaxies and galaxy clusters as well as superclusters.
For two first and two last size ranges are taken from the literature. 
For this paper we analyse the alignment of galaxies in the galaxy groups. 
As a sample 18 Tully's groups of galaxies \cite{Tully88} is taken. 

We obtained that there is no alignment of the Tully's groups. We interprete
this as a zero value of total angular momentum.
Taking into account that the $J\ne 0$ for the other sizes  we find that  
all five groups fits well to generalized Li relation.
It seems that there is simple relation $J(M)$ for the whole mass range while
the global rotation (even it is controversial) allowes us to obtain the
$J(M)$ relation which well fit to data.

This paper is organised in the following manner. In Section 2 for the
completence we present
 theoretical considerations, while Section 3 gives the main results
together with a discussion of the previous studies.
Section 4 concludes  the paper.

\section{Theoretical Considerations}

In homogeneous and isotropic models, the Universe with matter may not
only expand, but also rotate relative to the local gyroscope. The
motion of the matter can be described by the Raychaudhuri equation. This is
a relation between the scalar equation $\Theta$, the rotation tensor
$\omega_{ab}$ and the shear tensor $\sigma_{ab}$ \cite{Li98}.
The perfect fluid has the stress-energy tensor:
$T_{ab}=(\rho+p)u_au_b+pg_{ab}$,
where $\rho$ is the mass density and $p$ is the pressure. The Raychaudhuri
equation can be written as:
    $$
-\nabla_a A^a+\dot{\Theta}+{1\over3}\Theta^2+2(\sigma^2-\omega^2)=
{-4\pi G\over c^2}(\rho+3p),  \eqno(1)
    $$
where :
$A^a=u^b\nabla_bu^a$  is the acceleration vector, while
$\omega^2\equiv\omega_{ab}\omega^{ab}/2$ and
$\sigma^2\equiv\sigma^{ab}\sigma_{ab}/2$ are the scalar of rotation and shear
respectively.

It has been shown that a spatially homogeneous, rotating  and  expanding
Universe filled with a perfect fluid must have non-vanishing  shear
\cite{King73}.
 
Because $\sigma$ falls off  more  rapidly  than  the  rotation $\omega$ as the
Universe expands, it is reasonable  to  consider  such  generalization  of the
Friedman equation in which only the "centrifugal" term is present, i.e.
      $$
{\dot{a} ^2\over 2} +{\omega^2a^2\over 2} -{4\pi Ga^2\over 3c^2}\epsilon=
 - {kc^2\over 2},  \eqno(2)
      $$
where $\epsilon=\rho c^2$, $k$ is the curvature constant, $a$ a scalar factor
and $\dot{a}\equiv{d\over dt}$
 
Equation (2) should be completed with the principle of the conservation of
energy-momentum tensor and angular momentum:
         $$
\dot{\epsilon}=-(\epsilon +p)\Theta, \qquad  \Theta\equiv 3{\dot{a} \over a} \eqno(3)
         $$
         $$
{p+\epsilon \over c^2} a^5 \omega =J      \eqno(4)
         $$
From that we can observe that if  $p=0$  (dust),  then  $\rho\propto a^{-3}$
and $\omega\propto a^{-2}$, while in general $\sigma$ falls as $a^{-3}$
\cite{Hawking69}. The law of momentum conservation should
be satisfied for  each kind of matter, and consequently angular velocity of
the Universe will evolve according to different  laws  in  different epochs.
Before  the
decoupling ($z=1000$), matter and radiation interact, but  after  decoupling
dust  and  radiation  evolve  separately  and  they  have their own   angular
velocities $\omega_d$ and $\omega_r$. Parameters $\omega$ and $\rho$ can be
written as $\omega=\omega_0 (1+z)^2$,
$\rho=\rho_{d0} (1+z)^3 +\rho_{r0}(1+z)^4$, with the latter being
is the total mass density of matter and radiation.

 The conservation of angular momentum of a structure
 relative to the gyroscopic frames in dust epochs gives:
     $$
J=kM^{5/3} -lM,                            \eqno(5)
     $$
where $k={2\over5}\left({3\over 4\pi\rho_{d0}}\right)^{2/3}\omega_0$,
$\rho_{do}$ is the density of (dust) matter in the present epoch,
and $l=\beta r_f^2(1+z_f)^2\omega_0$, $r_f$ is the radius of protostructure, and
$\beta$ is a parameter determined by the distribution of the mass in it.
In this model we assumed that dark matter is collisionless and we considered
only barionic matter.

In \cite{Li98}  the present value of the angular velocity of the Universe
is estimated. A suitable value for $k$ parameter is assumed as 0.4 (in CGS Units).
 Taking  $\rho_{d0}= 1.88\cdot 10^{-29} \Omega h^2g\, cm^{-3}$
and $h=0.75$, $\Omega=0.01$
\cite{Peebles93} for rich clusters of galaxies, see also
\cite{Peebles02, Lahav02}, we obtain
$\omega_0\simeq 6\cdot 10^{-21}rad\, s^{-1} \simeq 2\cdot 10^{-13}rad\, yr^{-1}$

In our previous paper \cite{Godlowski03}
we show that relation (5) exhibits
a clear minimum (see Fig.1). The value of mass of such structure is:
      $$
M_{min}=\left({3l\over 5k}\right)^{3/2}=1.95 r_f^3 (1+z_f)^3 \rho_{d0}, \eqno(6)
      $$
and it does not depend on the present value of $\omega_0$ rotation of the
Universe.  We should note that when $|J|=0$, the corresponding mass
$M_{0}=\left({l\over k}\right)^{3/2} \approx 2.15 M_{min}$.
It is easy to observed that for less and more masive
structures considered model predict $|J(M)| \neq 0$. It is interesting that
from the observational point of view it is possible to observe just the cases
when $|J| \approx 0$. It means that  Li model can be verified by observations.

There  is a large number of scenario of galaxy cluster formation (see for
example \cite{Shandarin74, Wesson82, Silk83}).
We assumed that galaxy clusters are formed through the collapse of the
protostructures \cite{Sunyaev72, Doroshkevich73}.
For a protostructure with a diameter of $60 Mpc$ and $z_f=6$ , we obtain
$M_0 \approx 5\cdot 10^{13} {\cal M}_{\odot}$ as an estimate of dust mass for
a formed structure with vanishing angular momentum. This gives the total mass
of a structure of the order of
$10^{14} - 10^{15} {\cal M}_{\odot}$, which is a typical mass of a small
galaxy cluster. The Fig. 1 shows dependence of $J(M)$ in that case.

\begin{figure}
\vskip 10cm
\includegraphics{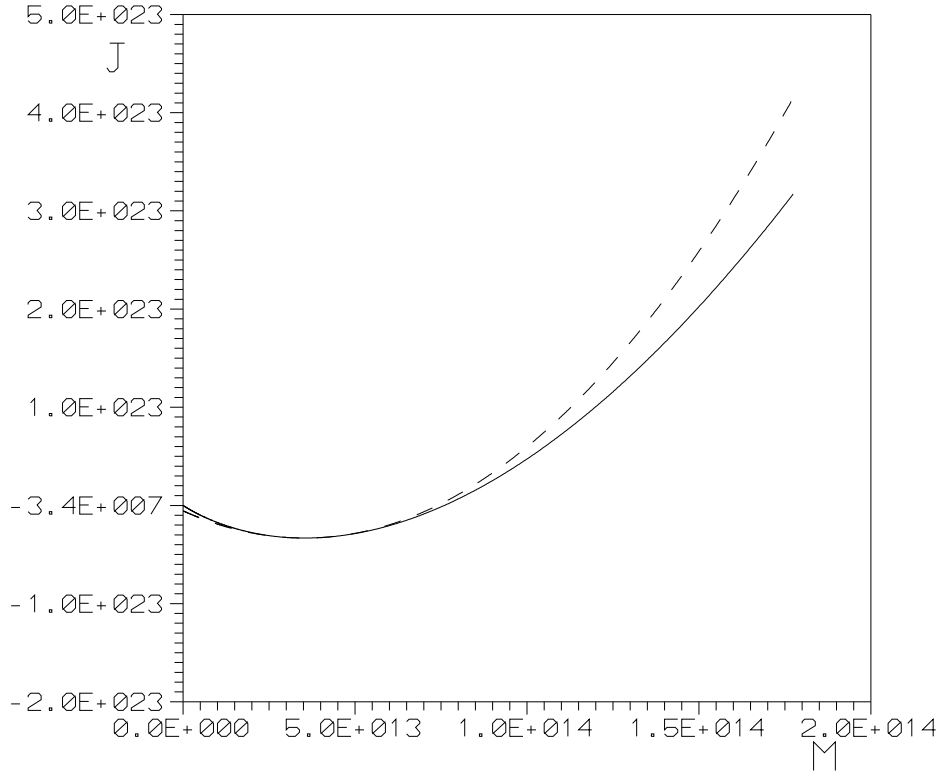}
\caption{The relation between angular momentum $J$ (in CGS units devided
by $10^{60}$) and M (in ${\cal M}_{\odot}$) of the astronomical object for
the protostructure with a diameter of $60 Mpc$ (solid line). Note that in
neighbourhood of minimum $J(M)$ relation Li approximated by
$J\propto M^{5/3}$ (dashes line).}
\end{figure}
 
\section{Results and Discussion}
 
The masses of the order $10^{14} - 10^{15} {\cal M}_{\odot}$, for which the
expected total angular momenta vanishes, are typical masses of the galaxy
groups. Therefore, we study the alignment of galaxies with the individual 
Tully's group in the LSC \cite{Tully88}.
So for the present study we chose 18 Tully's galaxy groups, each containing
more than 40 galaxies (see also \cite{Godlowski99}).
We investigated the galaxy alignment in each group
because the existence of alignment is interpreted as existence of
non-vanishing angular momentum
 
In our
analysis, we used the supergalactic coordinate system ($L$, $B$, $P$)
with the basic great circle (`meridian') chosen to pass through the LSC
centre in the Virgo cluster
\cite{Flin86, Godlowski93,  Godlowski94}
For any galaxy we consider two parameters:
the galactic position angle $p$ and the inclination angle $i$. With the
use of these angles two orientation angles are determined: $\delta$ -
angle between the normal to the galaxy and the LSC plane, and $\eta$ -
 angle between the projection of this normal on the the LSC plane and the
direction toward the LSC centre (seen from the Earth).
The distributions of the "supergalatic position angles" $P$, as well as two
angles $\delta$ and $\eta$ can be analyzed using statistical tests
briefly summarized in the Appendix. A detailed description
of our method can be found in \cite{Godlowski93, Godlowski94}.
 
The results of these statistical analyses are given in  Tables 1-3.
The tables show the results of  statistical tests
($\chi^2$ test, Fourier test and autocorrelation test).
Table 1 and 2 contains the results of the investigations of  galaxy plane
orientations.
It follows that a weak alignment of galaxy planes is observed.
However, the real alignment is very small, because a part of these
positive-signal detections is due to the observational effect in determining
galaxy axies in the Tully catalogue \cite{Godlowski99}.
Moreover, when we analysed spiral galaxies only, we found no group exhibiting
clear evidence for alignment existence.
The analysis of the distribution of galaxy position angles (Table 3)
shows non-randomness in one group only. So we conclude that in the case
of the Tully's group we do not find any galaxy alignment.

It should be noted that (because of the small number of object)
we repeated the derivations for different
values of $n$ bins (see Appendix), but no significant
differences were observed.
 
\begin{table}
\noindent
\caption{Test for isotropy of the orientations of galaxy plane. The distribution
of the angle $\delta$ of galaxies (Tully's group)}
\begin{tabular}{ccrrrrrrr}
angle&group&$N$&$\chi^2$&$C$&$P(\Delta_1)$&$\Delta_{11}$&$\sigma$\\
        & $11$ & 626&  62.8&   9.50& .000& -.237& .058\\
        & $12$ & 332&  25.7& -11.48& .299& -.125& .080\\
        & $13$ & 128&  29.7&  -6.94& .891& 0.002& .129\\
        & $14$ & 426&  24.0&   3.52& .154& -.120& .071\\
        & $15$ & 130&  13.1&  -1.88& .737& -.004& .128\\
        & $17$ &  80&  13.8&  -5.50& .569& -.100& .164\\
        & $21$ & 248&  14.2&   0.24& .065& -.035& .093\\
        & $22$ & 126&   7.0&   0.68& .496& 0.098& .130\\
$\delta$& $23$ & 100&  17.0&  -1.82& .607& 0.140& .146\\
        & $31$ & 210&  33.7&  15.98& .000& 0.137& .101\\
        & $41$ & 192&  22.8&   6.54& .020& -.004& .106\\
        & $42$ & 230&  24.7&  -3.78& .320& 0.056& .096\\
        & $44$ &  80&  26.7&   7.95& .024& -.200& .164\\
        & $51$ & 228&  29.6&  -0.41& .009& 0.093& .097\\
        & $52$ & 172&  21.6&   3.04& .005& -.203& .112\\
        & $53$ & 260&  13.3&  -3.29& .492& -.055& .091\\
        & $61$ & 258&  19.9&  -3.13& .825& -.049& .091\\
        & $64$ & 102&  28.7&  -7.54& .113& 0.080& .145\\
\end{tabular}
\end{table}
 
\begin{table}
\noindent
\caption{Test for isotropy of the orientations of galaxy plane. The distribution
of the angle $\eta$ of galaxies (Tully's group)}
\begin{tabular}{ccrrrrrrr}
angle&group&$N$&$\chi^2$&$C$&$P(\Delta_1)$&$\Delta_{11}$&$\sigma$\\
        & $11$ & 626&  60.0&   5.96& .000& 0.304& .057\\
	& $12$ & 332&  28.5&   7.56& .001& -.069& .078\\
        & $13$ & 128&  25.6&   2.78& .079& 0.242& .125\\
        & $14$ & 426&  27.4&   6.63& .090& 0.055& .069\\
        & $15$ & 130&  22.9&   2.51& .764& 0.036& .124\\
        & $17$ &  80&  13.1&  -5.97& .470& -.059& .158\\
        & $21$ & 248&  26.9&  -3.55& .054& 0.058& .090\\
        & $22$ & 126&  11.7&   5.57& .177& 0.194& .126\\
$\eta$  & $23$ & 100&  20.2&  -0.10& .081& -.164& .141\\
        & $31$ & 210&  24.0&   0.43& .046& 0.194& .098\\
        & $41$ & 192&  27.2&  10.22& .001& 0.300& .102\\
        & $42$ & 230&  15.9&   3.30& .033& 0.036& .093\\
        & $44$ &  80&  20.4&  -3.05& .226& 0.148& .158\\
        & $51$ & 228&  30.6&  -5.37& .042& 0.218& .094\\
        & $52$ & 172&  38.1&  12.29& .001& 0.212& .108\\
        & $53$ & 260&  12.2&  -8.28& .816& 0.008& .088\\
        & $61$ & 258&  23.7& -12.56& .549& 0.091& .088\\
        & $64$ & 102&  50.1&  -3.53& .002& 0.402& .140
\end{tabular}
\end{table}

\begin{table}
\noindent
\caption{Test for isotropy of the distribution of supergalactic position
angles P of galaxies (Tully's group)}
\begin{tabular}{ccrrrrrrr}
angle&group&$N$&$\chi^2$&$C$&$P(\Delta_1)$&$\Delta_{11}$&$\sigma$\\
    & $11$ &185 & 22.7& -11.42& .728&  0.081& .104\\
    & $12$ &106 & 17.6&  -1.57& .198&  -.083& .137\\
    & $13$ & 50 & 13.4&  -2.48& .714&  -.056& .200\\
    & $14$ &133 & 14.9&  -1.05& .990&  -.011& .123\\
    & $15$ & 48 & 11.3&  -0.75& .185&  -.006& .204\\
    & $17$ & 22 & 10.7&  -5.64& .727&  -.152& .302\\
    & $21$ & 85 & 13.5&  -2.84& .878&  -.058& .153\\
    & $22$ & 43 & 16.0&  -1.98& .910&  -.089& .216\\
$P$ & $23$ & 33 & 12.3&   0.27& .230&  -.337& .246\\
    & $31$ & 63 & 20.1&   1.00& .729&  0.081& .178\\
    & $41$ & 54 & 20.0&  10.67& .595&  -.112& .192\\
    & $42$ & 71 & 19.0&   1.51& .124&  -.254& .168\\
    & $44$ & 25 & 18.9&   1.64& .367&  0.161& .283\\
    & $51$ & 69 & 23.1&   3.00& .576&  -.176& .170\\
    & $52$ & 50 & 14.8&  -0.32& .631&  0.154& .200\\
    & $53$ & 88 & 17.5&   2.82& .080&  0.243& .151\\
    & $61$ & 85 & 27.9&   6.48& .007&  -.116& .153\\
    & $64$ & 31 & 18.4&   2.10& .295&  -.397& .254
\end{tabular}
\end{table}

There is no clear empirical evidence that galaxy groups and cluster rotate.
However, some evidence connected with the motion of their
components around the mass centre in very poor groups can be  observed.
As a result, it can be accepted that the angular momentum of a galaxy group or
cluster is connected mainly with the galaxy spins. Even when taking into
account the eventual orbital motion of galaxies around the common mass
centre, it can be easily seen that more numerous structures exhibit larger
angular momentum.
Because of that reason, we assumed that total angular momentum of the galactic
groups and clusters consists from the sum of spins of their components only.
For masses of structure close to $M_0$ we do not find evidence that
angular momenta $J(M) \neq 0$, which is in agreement with predictions of the
considered Li model.
 
The less massive structures are compact groups and galaxy pairs.
In compact groups of galaxies, member galaxies  rotate along elongated orbits
around the gravitational centre of the group  \cite{Tiersch01}
which obviously add some angular momentum to the total angular
momentum of the system.
 
The study of paired galaxies showed that the angular momentum in these
systems becomes mainly in the orbital motion  of galaxies
\cite{Karachentsev84a, Karachentsev84b, Mineva87}.  It
showed that for the structures with masses smaler then $M_0$ there are
some indicators of the non-vanishing angular momentum.
 
 The study of galaxy orientation, which substitutes the
investigation of the galaxy spin distribution, yields different results.
Nevertheless, it is clear that in isolated  Abell clusters of galaxies only the
dominant  brightest cluster members exhibit the sign of alignment 
\cite{Flin91, Trevese92,  Kim01}.
However, in very rich galaxy clusters, such  as A754 \cite{Godlowski98},
or A1656 \cite{Djorgovski83, Wu97, Wu98} the
non-random distribution of galaxies has been observed.

A lot of work has been done in the study of the alignment of galaxies in
superclusters (for earlier work, see the review of Djorgovski \cite{Djorgovski87}.
The results are ambigous, but several independent
investigations claimed the existence of a weak alignment of galaxies in
respect to the supercluster's main plane. This was  the case of the Local
Supercluster \cite{Flin86, Kashikawa92, Godlowski93, Godlowski94},
Hercules Supercluster \cite{Flin94},
Coma/A1367 \cite{Djorgovski83, Wu97, Wu98, Flin01},
and the Perseus Supercluster \cite{Flin88, Flin89a, Flin89b}.
In the latter supercluster, the weak alignment has been observed also when data
taken from scans, i.e. without a personal bias \cite{Cabanela98},
were involved instead of the  data derived from visual measurements.
The distribution of real spins was random, but the sample considered was
small  \cite{Cabanela99}. The above-mentioned studies were
restricted to the high-density regions of superclusters.
From the above considerations it follows that the alignment of galaxies in
Abell clusters is much weaker than in the case of superclusters.

Summarising our results, there is an evidence of non-vanishing angular momentum
in both small and larger structures, but not for galaxy groups.
 Our results also show that during analising our sample of the galaxy groups,
the most prominent possible alignment of galaxy plane orientation, we obtain
for the Virgo cluster itself, the most massive substructure studied by us.
This results are consistent with predictions of the discussed Li model.
 
\section{Conclusions}
 
 In the present paper we discuss the relation between angular momentum and
mass of the galaxy structures obtained from  Li model.
 The classical relation  $J \propto M^{5/3}$ implies a monotonical increase
of angular momentum with the mass. We show that this simple relation  is not
confirmed by observations. In the Li model simple relation $J\propto M^{5/3}$
is valid only for larger masses,
while the model predicts the presences of a minimum in the
relation beetwen angular momenta  and masses of galaxy structures.
It is interesting that the zero of $J(M)$ relation can be tested observationally.
We demonstrate that with resonable assumptions
the mass $M_0$ for which $|J|=0$, corresponds to galaxy group mass.

Analysing Tully's groups of galaxies we do not find any evidence for
galaxy alignment for that groups.
Our results - evidence for vanishing angular momentum for the
galaxy grups and indicators of non-vanishing angular momentum for less and
more masive structures is consistent with predictions of the discussed model.

Our general conclusion is that  Li model in which celestial bodies
acquire  angular momenta  during their formation from the
global rotation of the Universe, gives us correct predictions.
This for example explain why for galaxy groups, contrary to the more massive
structures, no alignment of galaxies is dedected.

\section*{Appendix. Statistical methods applied}

In order to check the distribution of galaxy orientation angles ($\delta,
\eta$) and position angles $p$, we tested whether the respective
 distribution of the $\delta, \eta$ or $p$ angles is isotropic.
We applied statistical tests originally
introduced (for that problem) by \cite{Hawley75, Kindl87},
but  modified by us as compared to the original version
(see the detailed description in \cite{Godlowski93, Godlowski94}.
Below, a short summary  is presented of the tests
considered here (not always explicitly): the $\chi^2$-test, the
Fourier test and the auto-correlation test.
In  all of these  tests,  the entire range of the $\theta$ angle
(where for $\theta$ one can put $\delta+\Pi/2$, $\eta$ or $p$ respectively)
is divided into $n$ bins, which in the $\chi^2$ test gives $n-1$
degrees of freedom. During the analysis,
we used  $n = 18$ bins of equal width.
Let $N$ denote the total number of galaxies in the considered cluster, and
$N_k$ - the number of galaxies with orientations within the $k$-th  angular
bin. Moreover,  $N_0$ - denotes the average number of  galaxies  per
bin and, finally, $N_{0,k}$ - the expected number of galaxies in the $k$-th
bin. The $\chi^2$-test of the distribution yields the critical value 27.6
(at the siginificance level $\alpha =0.05$) for 17  degrees  of freedom:
 
$$\chi^2 = \sum_{k = 1}^n {(N_k -N_{0,k})^2 \over N_{0,k}} \qquad .
\eqno(A1)$$
 
\noindent
 
However, when we consider individual clusters the number of galaxies
involved may be small in some cases, and the $\chi^2$ test will not neccesarily
 work well (e.g. the $\chi^2$
test requires the expected number of data per bin to equal at least 7;
see, however, \cite{Snedecor67, Domanski79}.
As a check, in a few cases we repeated the derivations for different
values of n, but no significant differences appeared.
However, the main statistical test used in the present paper is the Fourier test.
In the Fourier test the actual distribution $N_k$ is approximated as:
 
$$N_k = N_{0,k} (1+\Delta_{11} \cos{2 \theta_k} +\Delta_{21} \sin{2 \theta_k}
 \qquad ,   \eqno(A2)$$

\noindent
(we take into account only the first Fourier mode).
We obtain the following expression for the
coefficients $\Delta_{ij}$ ($i,j = 1, 2$):
 
$$\Delta_{1j} = {\sum_{k = 1}^n (N_k -N_{0,k})\cos{2J \theta_k} \over
\sum_{k = 1}^n N_{0,k} \cos^2{2J \theta_k} } \qquad , \eqno(A3)$$
 
$$\Delta_{2j} = { \sum_{k = 1}^n (N_k-N_{0,k})\sin{2J \theta_k} \over
\sum_{k = 1}^n N_{0,k} \sin^2{2J \theta_k} } \qquad , \eqno(B4)$$\\
 
\noindent
with the standard deviation
 
$$ \sigma(\Delta_{11}) = \left( {\sum_{k = 1}^n N_{0,k} \cos^2{2 \theta_k}
} \right)^{-1/2} \approx \left( {2 \over n N_0} \right)^{1/2} \qquad ,
\eqno(A5a)$$
 
$$ \sigma(\Delta_{21}) = \left( {\sum_{k = 1}^n N_{0,k} \sin^2{2 \theta_k}
} \right)^{-1/2} \approx \left( {2 \over n N_0} \right)^{1/2} \qquad .
\eqno(A5b)$$
 
\noindent
where $N_0$ is the average of all $N_{0,k}$.
However, we should note that we could formally replace the symbol $\approx$
with $=$ only in the cases where all $N_{0,k}$ are equal (for example, in the
cases when we tested the isotropy of the distribution of the position angle).
The probability that the amplitude:
 
$$\Delta_1 = \left( \Delta_{11}^2 + \Delta_{21}^2 \right)^{1/2}
\eqno(A6)$$
 
\noindent
is greater than a certain chosen value is given by the formula:
 
$$P(>\Delta_1 ) = \exp{\left( -{n \over 4} N_0 \Delta_1^2 \right)}
\qquad , \eqno(A7)$$
 
\noindent
while the standard deviation of this amplitude is
 
$$ \sigma(\Delta_1) = \left( {2 \over n N_0} \right)^{1/2} \qquad .
\eqno(A8)$$
 
From the value of $\Delta_{11}$ one can deduce the direction of the
departure from isotropy. If $\Delta_{11} < 0$, then, for $\theta \equiv
\delta + \pi/2$, an excess of galaxies with rotation axes parallel to
the LSC plane is present. For $\Delta_{11} > 0$ the rotation axes
tend to be perpendicular to the LSC plane.
Similarly, while analysing the distribution of the position angles of galaxies
($\theta \equiv p$), if $\Delta_{11} < 0$, an excess of galaxies with
position angles parallel to the plane of the coordinate system (i.e. normal 
to the galaxy plane is  perpendicular to the plane of the coordinate system)
is present. For $\Delta_{11} > 0$, the position angles of galaxy are 
perpendicular to the plane of the coordinate system.
 
The auto-correlation test quantifies the correlations between the
galactic numbers in adjoining angular bins. The correlation function is
defined as:
 
$$C\, = \, \sum_{k = 1}^n { (N_k -N_{0,k})(N_{k+1} -N_{0,k+1} )
 \over \left[ N_{0,k} N_{0,k+1}\right]^{1/2} } \qquad . \eqno(A9)$$
 
\noindent
In the case of an isotropic distribution we expected $C = 0$ with the standard
deviation:
 
$$\sigma(C) = n^{1/2} \qquad . \eqno(A10)$$

\end{document}